\documentclass[prd,showkeys,floatfix,nofootinbib,
               preprint,12pt,
               tightenlines,fleqn]{revtex4-1} 

\usepackage{amsmath,amssymb,revsymb,graphicx,dcolumn}
\usepackage{leftidx}
\usepackage{slashed}
\usepackage{marginnote}

\usepackage{xcolor}
\definecolor{mypink}{rgb}{0.9, 0.1, 0.3}

\newcommand{\beq}{\begin{equation}}
\newcommand{\eeq}{\end{equation}}
\newcommand{\beqa}{\begin{eqnarray}}
\newcommand{\eeqa}{\end{eqnarray}}
\newcommand{\bsubeqs}{\begin{subequations}}
\newcommand{\esubeqs}{\end{subequations}}

\DeclareMathOperator{\csch}{csch}

\begin{document}
\title[]
      {On quantum corrections to geodesics in de-Sitter spacetime}
\author{Viacheslav A. Emelyanov}
\email{viacheslav.emelyanov@kit.edu}
\affiliation{Institute for Theoretical Physics,\\
Karlsruhe Institute of Technology,\\
76131 Karlsruhe, Germany\\}

\begin{abstract}
\vspace*{2.5mm}\noindent
We find a coordinate-independent wave-packet solution of the massive Klein-Gordon equation
with the conformal coupling to gravity in the de-Sitter universe. This solution can locally be
represented through the superposition of positive-frequency plane waves at any space-time point,
assuming that the scalar-field mass $M$ is much bigger than the de-Sitter Hubble constant $H$.
The solution is also shown to be related to the two-point function in the de-Sitter quantum
vacuum. Moreover, we~study the wave-packet propagation over cosmological times, depending on the
ratio of $M$ and $H$.~In~doing so, we find that this wave packet propagates like a point-like particle of
the same mass if $M \ggg H$, but, if otherwise, the wave packet behaves highly non-classically.
\end{abstract}


\maketitle

\section{Introduction}

Elementary particles in Minkowski spacetime are related to unitary and irreducible
representations of the Poincar\'{e} group. Their notion is thus unambiguous in all
Lorentz frames, as the Poincar\'{e} group represents the isometry group of
Minkowski spacetime. It might~be~then tempting to expect that there is no
well-defined vacuum notion in curved~spacetimes.~In~fact, 
Schr\"{o}dinger argued that particles may be produced in
evolving universes~\cite{Schroedinger}.
This quantum effect arises from the absence of time-translation symmetry, which requires
the re-definition of creation and annihilation
operators during time evolution, while quantum states remain
unchanged. A no-particle state at earlier times
may not then be interpreted as an empty state at later times~\cite{Parker1969,Parker&Toms}. In addition,
a single-particle state turns into a multi-particle~state~over time,
resembling, thereby, particle decays in interacting quantum-field models.

In spite of the fact that the observable Universe is dynamically changing all the time,
we successfully describe high-energy processes by using the Standard Model of
particle physics, in which the Poincar\'{e} group plays a crucial role~\cite{Weinberg}.
In the Standard Model, a particle decay may occur if compatible
with various conservation laws. In particular, we observe on Earth
that energy, momentum and angular momentum are conserved in particle scatterings.
As an example, the electron neutrino was foreseen in $\beta$-decay from
energy-momentum conservation long before its actual detection~\cite{Pauli}.
These conservation laws are, in turn, related to the space-time translation and
rotational symmetries which are spontaneously broken in nature. Still, these laws must
locally hold, according to the equivalence principle, which is in agreement
with the up-to-date observations~\cite{Will}.

It is thus an empirical fact that particles in collider physics are well defined, even
though the observable Universe is evolving. This could be readily explained if
wave functions, which describe particles, are well localised in spacetime. Their nonpoint-like
support is still testable in gravity, namely the quantum interference of non-relativistic
neutrons was observed in the Earth's gravitational field~\cite{Colella&Overhauser&Werner}.
This observation is consistent with the Schr\"{o}dinger equation with the Newtonian
potential. In general, if the Compton wavelength of particles is negligible
with respect to a characteristic curvature length, then the quantum
interference induced by gravity can be described by the coordinate-independent
(covariant) phase factor
\beqa\label{eq:covariant-phase-factor}
\exp\Big({-}iM{\int_A^B}ds\Big) \quad \text{with} \quad
ds^2 \;=\; g_{\mu\nu}(x)\,dx^\mu dx^\nu\,,
\eeqa
where $M > 0$ is the particle mass, $A$ and $B$ are, respectively,
initial and final positions of~the particle which moves along a geodesic
connecting these points~\cite{Stodolsky}.

The main purpose of this article is to generalise this result to the case when the Compton
wavelength of particles may be comparable with the characteristic curvature length. Besides,
particle's propagation time may be as large as a characteristic curvature time. Since
it~is~not obvious if this generalisation is even possible, we shall consider de-Sitter spacetime, in
which one should be able to find a non-perturbative result due to de-Sitter symmetries.

Throughout, we use natural units $c = G = \hbar = 1$, unless otherwise stated.

\section{Adiabatic particles in de-Sitter spacetime}
\label{sec:cpd-ds}

It was elaborated in 1968 how adiabatic particles may be created in an expanding universe
in linear quantum field models~\cite{Parker1969}. In this section, we briefly review this
adiabatic-particle-creation process in the de-Sitter universe in order to introduce concepts and
notations which will be used later on.

Considering de-Sitter spacetime with the Hubble parameter $H$ in flat coordinates $(t,\mathbf{x})$,
i.e. the spatial curvature equals zero in these coordinates, the particle-creation operator can be defined
through the adiabatic modes at past and future cosmic infinities~\cite{Parker&Toms}.
Following~the recent references~\cite{Anderson&Mottola,Anderson&Mottola&Sanders}, one has
\beqa\label{eq:pco}
\hat{a}^\dagger(\varphi_\mathbf{k}) &=& \left\{
\begin{array}{llcc}
\hat{a}^\dagger(\varphi_{\mathbf{k},-\infty})\,, & t &\rightarrow& -\infty\,, \\[2mm]
\hat{a}^\dagger(\varphi_{\mathbf{k},+\infty})\,, & t &\rightarrow& +\infty\,,
\end{array}
\right.
\eeqa
where, in case of the scalar field $\Phi(x)$ with the mass $M$ and conformal coupling to gravity,
\bsubeqs\label{eq:a-modes}
\beqa\label{eq:a-modes-in}
\varphi_{\mathbf{k},-\infty}(x) &=& \bigg(\frac{\pi}{4Ha^3(t)}\bigg)^\frac{1}{2}e^{\frac{\pi i}{4}-\frac{\pi\mu}{2}}
H_{i\mu}^{(1)}\bigg(\frac{|\mathbf{k}|}{Ha(t)}\bigg)\,e^{i\mathbf{k}\mathbf{x}}\,,
\\[1mm]\label{eq:a-modes-out}
\varphi_{\mathbf{k},+\infty}(x) &=& \bigg(\frac{1}{2\mu Ha^3(t)}\bigg)^\frac{1}{2}2^{i\mu}\,\Gamma(1+i\mu)\,
J_{i\mu}\bigg(\frac{|\mathbf{k}|}{Ha(t)}\bigg)\,e^{i\mathbf{k}\mathbf{x}}\,,
\eeqa
\esubeqs
where $a(t) = e^{Ht}$ is the de-Sitter scale factor, $\Gamma(z)$,
$H_{i\mu}^{(1)}(z)$ and $J_{i\mu}(z)$ are, respectively, the gamma, Hankel and Bessel functions, and
\beqa
\mu &\equiv& \frac{1}{2}\sqrt{4\nu^2-1} \; >\; 0 \quad \text{with} \quad \nu \;\equiv\; M/H\,.
\eeqa
The exact solutions $\varphi_{\mathbf{k},-\infty}(x)$ and $\varphi_{\mathbf{k},+\infty}(x)$ match, respectively,
the adiabatic modes at past and future infinity. These will be referred to as the past
and future adiabatic modes.

The de-Sitter universe turns into Minkowski spacetime in the limit $H \rightarrow 0$. It
is straightforward to show in this case that both asymptotic adiabatic modes
turn into the Minkowski plane-wave solutions up to a phase factor. However,
$\varphi_{\mathbf{k},-\infty}(x)$ gives rise to a ``preferred"
state in de-Sitter spacetime. In fact, this mode defines
the Chernikov-Tagirov aka Bunch-Davies state~\cite{Chernikov&Tagirov,Bunch&Davies},
which we denote by $|\text{dS}\rangle$.

This quantum state is a no-adiabatic-particle state at past infinity (in flat de-Sitter space),
in the sense that $|\text{dS}\rangle$ is annihilated by $\hat{a}(\varphi_{\mathbf{k},-\infty})$, i.e.
$\hat{a}(\varphi_{\mathbf{k},-\infty})|\text{dS}\rangle = 0$ and
\beqa
\hat{a}(\varphi_{\mathbf{k},-\infty}) &\equiv&
+i{\int_\Sigma}d\Sigma^{\mu}(x)
\big(\overline{\varphi}_{\mathbf{k},-\infty}(x)\nabla_\mu\hat{\Phi}(x)
- \hat{\Phi}(x)\nabla_\mu\overline{\varphi}_{\mathbf{k},-\infty}(x)\big)\,,
\eeqa
where $\Sigma$ is a space-like Cauchy surface and the bar stands here for
the complex conjugation. A normalisable single-$\varphi_{-\infty}$-particle state can
be then defined as
\beqa
|\varphi_{f_\mathbf{p},-\infty}\rangle &\equiv& {\int}\frac{d^3\mathbf{k}}{(2\pi)^3}\,
f_\mathbf{p}(\mathbf{k})\,\hat{a}^\dagger(\varphi_{\mathbf{k},-\infty})|\text{dS}\rangle
\;\equiv\; \hat{a}^\dagger(\varphi_{f_\mathbf{p},-\infty})|\text{dS}\rangle\,,
\eeqa
where $f_\mathbf{p}(\mathbf{k})$ is a square-integrable function sharply peaked at
$\mathbf{k} = \mathbf{p}$, such that
\beqa
\langle \varphi_{f_\mathbf{p},-\infty}|\varphi_{f_\mathbf{p},-\infty}\rangle &=&
{\int}\frac{d^3\mathbf{k}}{(2\pi)^3}\,
|f_\mathbf{p}(\mathbf{k})|^2 \;\equiv\; 1\,.
\eeqa
The state $|\varphi_{f_\mathbf{p},-\infty}\rangle$ does not depend on $t$, in
accordance with the Heisenberg picture we have been working in.
Therefore, $|\text{dS}\rangle$ is empty with respect to
$\hat{a}(\varphi_{\mathbf{k},-\infty})$ at all time moments and the de-Sitter particles
are related to unitary and irreducible representations of the de-Sitter symmetry
group~\cite{Nachtmann1967}. These particles may be dynamical, i.e.
$|\varphi_{f_\mathbf{p},-\infty}\rangle$ depends on time,~only
in interacting field models~\cite{Nachtmann1968,Bros&Epstein&Moschella}.

From another side, the de-Sitter mode $\varphi_{\mathbf{k},-\infty}(x)$ turns into a linear
superposition of the positive- and negative-frequency adiabatic modes at future infinity:
\beqa\label{eq:btwp}
\varphi_{\mathbf{k},-\infty}(x) &=& \alpha(\varphi_{+\infty},\varphi_{-\infty})
\,\varphi_{\mathbf{k},+\infty}(x) +
\overline{\beta}(\varphi_{+\infty},\varphi_{-\infty})\,\overline{\varphi}_{\mathbf{k},+\infty}(x)\,,
\eeqa
where the Bogolyubov coefficients can be found
in~\cite{Anderson&Mottola,Anderson&Mottola&Sanders}. This leads to
\beqa\label{eq:bto}
\hat{a}^\dagger(\varphi_{\mathbf{k},-\infty}) &=& \alpha(\varphi_{+\infty},\varphi_{-\infty})
\,\hat{a}^\dagger(\varphi_{\mathbf{k},+\infty})
- \overline{\beta}(\varphi_{+\infty},\varphi_{-\infty})\,\hat{a}(\varphi_{-\mathbf{k},+\infty})\,.
\eeqa
Hence, the adiabatic-particle-number operator
$\hat{N}(\varphi_{f_\mathbf{p}}) = \hat{a}^\dagger(\varphi_{f_\mathbf{p}})\hat{a}(\varphi_{f_\mathbf{p}})$
changes with cosmic time. In particular, one has
\beqa
\langle\text{dS}|\hat{N}(\varphi_{f_{\pm\mathbf{p}}}) |\text{dS}\rangle &=& \left\{
\begin{array}{llcc}
0\,, & t &\rightarrow& -\infty\,, \\[2mm]
|\beta(\varphi_{+\infty},\varphi_{-\infty})|^2\,, & t &\rightarrow& +\infty\,.
\end{array}
\right.
\eeqa
This means that $|\text{dS}\rangle$ is a $N$-adiabatic-particle state at future infinity, assuming that
\beqa
N \equiv \text{floor}\big(|\beta(\varphi_{+\infty},\varphi_{-\infty})|^2\big)\,,
\eeqa
where $N$ can be arbitrarily large, as the
Pauli principle does not apply to bosons. This~result is known in the literature as the cosmological
adiabatic-particle creation~\cite{Parker1969,Parker&Toms}.

This particle creation is based on the re-definition of the particle notion
over time (see~\eqref{eq:pco}). The procedure implies that adiabatic
particles are unstable. Specifically,~if~$|\varphi_{f_\mathbf{p},-\infty}\rangle$~describes
a single-adiabatic-particle state at past infinity, then this state should be
re-interpreted~as~a multi-adiabatic-particle state at future infinity. In fact, one finds that
\bsubeqs
\beqa
\langle\varphi_{f_\mathbf{p},-\infty}|\hat{N}(\varphi_{f_{+\mathbf{p}},+\infty})|\varphi_{f_\mathbf{p},-\infty}\rangle
&=& 1 + 2|\beta(\varphi_{+\infty},\varphi_{-\infty})|^2\,,
\\[2mm]
\langle\varphi_{f_\mathbf{p},-\infty}|\hat{N}(\varphi_{f_{-\mathbf{p}},+\infty})|\varphi_{f_\mathbf{p},-\infty}\rangle
&=& 2|\beta(\varphi_{+\infty},\varphi_{-\infty})|^2\,.
\eeqa
\esubeqs
This result may be called as the cosmological adiabatic-particle decay.

A particle decay in interacting field models is the process that may take place
if it does not violate various conservation laws. For instance, we observe on Earth
that energy, momentum and angular momentum are conserved in collider physics.
These conservation laws come from
space-time translation and rotational symmetries which are local symmetries 
of the Universe, according to the equivalence principle. In contrast,
the cosmological particle decay cannot~be a local process: The gravitational field is
the only source of energy which is available for~this decay, but the gravitational-field energy is
non-localisable~\cite{Misner&Thorne&Wheeler}.

\section{Covariant particles in de-Sitter spacetime}
\label{sec:cpp-cs}

\subsection{Motivation}

A scattering process in particle physics usually corresponds to unitary evolution of
a $N$-particle state defined at past infinity into a $N$-particle state defined at
future infinity. It is evident though that it is impossible to carry out a scattering experiment
with the asymptotic states, i.e. states defined at $t \rightarrow \pm\infty$, in collider physics,
bearing in mind that initial states should then have been arranged at the Big Bang. This apparent
tension between theoretical constructions and experiments can be eliminated by taking
into account that elementary particles are quantum-field excitations localised in spacetime, namely
they are described by wave packets with a finite space-time extent. A Wilson cloud
chamber is actually designed to visualise a charged-particle trajectory which is localised within the
chamber and, hence, in space. Besides, the free neutron decays into a proton, electron and
electron antineutrino, with a mean lifetime of around $10^3$ seconds -- free neutrons
are also localised in time. For these reasons, initial/final $N$-particle states need to be arranged
not at past/future infinity, but rather at a fraction of a second before/after the scattering process.
This means particles are essentially non-interacting if their wave packets are
well-separated. This observation also explains why the Minkowski-spacetime approximation
used in theory works well in practice: The observable Universe locally looks as Minkowski
spacetime and, consequently, particles can be considered within
a local inertial frame, since their support is normally much smaller than the local-frame extent.

The question of our interest is how the asymptotic states of collider physics emerge locally
in curved spacetime. These quantum states describe elementary particles which are free of
interactions. In the field model under consideration, this means that we need to determine
a single-particle state which can describe a scalar particle to move along a geodesic.

One of the fundamental properties of the geodesic equation is its form invariance under general coordinate
transformations. For example, geodesics do not depend on the coordinate parametrisation of
de-Sitter spacetime. However, in the closed coordinates to cover the entire de-Sitter
hyperboloid, adiabatic modes at past time infinity and the de-Sitter modes do not
match~\cite{Anderson&Mottola}. The notion of an adiabatic particle is, in general,
coordinate-dependent.

Another basic property of geodesics is that they locally reduce to straight lines. That is
a free-particle trajectory $x(\tau)$, where $\tau$ is the proper time, is locally
of the form $x(0) + \dot{x}(0)\,\tau$, where $x(0)$ and $\dot{x}(0)$ are the
particle position and velocity at $\tau = 0$, respectively. In quantum theory over Minkowski spacetime,
a constant-momentum single-particle state is described by the plane-wave-mode superposition.
According to the equivalence principle, this description must also hold in a
Fermi normal frame related to a particle geodesic in de-Sitter spacetime
if $H|\Delta{t}| \ll 1$ and $H|\Delta\mathbf{x}| \ll 1$, where $|\Delta{t}| = |\Delta\mathbf{x}| = 0$
corresponds to that geodesic. In particle physics, we have
also to require that $H\lambda_c \lll 1$, where $\lambda_c$ is the Compton wavelength of
the elementary particle (see below). Under these premises, quantum field theory over Minkowski
spacetime should adequately describe this particle locally.

As noted above,
$\varphi_{\mathbf{k},-\infty}(x)$ turns into the Minkowski plane-wave solution if
$H \rightarrow 0$. Since the Hubble parameter is dimensionful, one needs
instead to consider $H \ll M$ and $H|t| \ll 1$.
The first condition is fulfilled by the massive fields of the Standard Model if $H$ is identified
with the present Hubble parameter, $H_0 \sim 10^{-26}\,\text{m}^{-1}$. The second condition
cannot hold for all times. It is known by now that the dark-energy-dominated epoch has
started at around $10^{16}\,\text{s}$ after the Big Bang, whereas the universe age
$t_0 \sim 1/H_0$ is about $10^{18}\,\text{s}$~\cite{Mukhanov}.
Therefore, $\varphi_{\mathbf{k},-\infty}(x)$ cannot be reduced to the plane-wave mode all
the time over the present de-Sitter-like epoch. From another side,
plane-wave modes are successfully applied in particle physics to describe
high-energy scattering processes which were taking place over the entire semi-classical
history of the Universe.

The later circumstance shows that neither $\varphi_{\mathbf{k},-\infty}(x)$ nor
$\varphi_{\mathbf{k},+\infty}(x)$ are appropriate for our goal. We intend below to derive a
covariant wave-packet solution of the scalar-field~equation,
\beqa\label{eq:sfe}
\Big(\Box_x + M^2 - \frac{1}{6}\,R(x)\Big)\Phi(x) &=& 0 
\quad \text{with} \quad
R(x) \;=\; -12H^2\,,
\eeqa
which can be locally represented through the superposition of positive-frequency
plane waves at any space-time point.

\subsection{Covariant wave packet in Minkowski spacetime}
\label{sec:ms:cwp}

In particle physics in Minkowski spacetime, a particle, which is localised
at $X = (T,\mathbf{X})$ in position space and at $P = (P^T,\mathbf{P})$
in momentum space (with localisation regions in both spaces related
through the uncertainty relation~\cite{Merzbacher}), is described by the state
\beqa
|\varphi_{X,P}\rangle &\equiv& {\int}\frac{d^4K}{(2\pi)^3}\,\theta(K^T)\,\delta(K^2-M^2)\,
F_{P}(K)\,e^{+iK{\cdot}X}\,\hat{a}^\dagger(\mathbf{K})|\text{M}\rangle\,,
\eeqa
where $F_{P}(K)$ is peaked at $K = P$ and the state
$|\text{M}\rangle$ stands here for the Minkowski quantum vacuum.
The particle-creation operator in momentum
space is
\beqa\label{eq:ms:msco}
\hat{a}^\dagger(\mathbf{K}) &\equiv& - i{\int_t}d^3\mathbf{x}\,
\big(e^{-iK{\cdot}x}\partial_t \hat{\Phi}(x) - \hat{\Phi}(x)\partial_te^{-iK{\cdot}x}\big)\,,
\eeqa
which satisfies the commutation relation
$[\hat{a}(\mathbf{K}),\hat{a}^\dagger(\mathbf{P})] =
2\sqrt{\mathbf{K}^2 + M^2}\,(2\pi)^3\delta(\mathbf{K}-\mathbf{P})$. This
straightforwardly follows from the commutator of the scalar-field operator at different
space-time points. The operators $\hat{a}^\dagger(\mathbf{K})$ and $\hat{a}(\mathbf{K})$ provide
the standard expansion of the quantum field $\hat{\Phi}(x)$ over the creation and annihilation
operators.

The function $F_{P}(K)$ is chosen in such a way that the state
$|\varphi_{X,P}\rangle$ is normalised to unity:
\beqa
\langle \varphi_{X,P}|\varphi_{X,P}\rangle &=& \frac{1}{2}{\int}\frac{d^3\mathbf{K}}{(2\pi)^3}\,
\frac{|F_{P}(\mathbf{K})|^2}{\sqrt{\mathbf{K}^2 + M^2}} \;\equiv\; 1\,.
\eeqa
We refer to the reference~\cite{Itzykson&Zuber} for further details.

\subsubsection{Gaussian wave packet in Minkowski spacetime}

For later applications, however, it proves useful to introduce a wave packet describing the
particle state $|\varphi_{X,P}\rangle$. Specifically, this wave packet reads
\beqa\label{eq:lwp-ms}
\varphi_{X,P}(x) &\equiv& {\int}\frac{d^4K}{(2\pi)^3}\,\theta(K^T)\,\delta(K^2-M^2)\,
F_{P}(K)\,e^{-iK{\cdot}(x - X)}\,,
\eeqa
giving rise to
\beqa
\hat{a}^\dagger(\varphi_{X,P}) &=& 
- i{\int_t}d^3\bold{x}\,
\big(\varphi_{X,P}(x)\partial_t \hat{\Phi}(x) - \hat{\Phi}(x)\partial_t\varphi_{X,P}(x)\big)\,,
\eeqa
which produces the state $|\varphi_{X,P}\rangle = \hat{a}^\dagger(\varphi_{X,P})|\text{M}\rangle$.
The normalisation condition in terms~of~the wave packet $\varphi_{X,P}(x)$ takes
the form
\beqa
- i{\int_t}d^3\bold{x}\,
\big(\varphi_{X,P}(x)\partial_t \overline{\varphi}_{X,P}(x)
- \overline{\varphi}_{X,P}(x)\partial_t\varphi_{X,P}(x)\big) &=& 1\,
\eeqa
(cf.~(6) and the text below that equation in the reference~\cite{Schroedinger}).

In the absence of self-interaction or interaction with other quantum fields,
$\hat{a}^\dagger(\varphi_{X,P})$ must be time-independent. This is realised if
the wave packet vanishes sufficiently fast in the limit
$|\mathbf{x}-\mathbf{X}| \rightarrow \infty$. Considering a
Lorentz-invariant Gaussian wave packet~\cite{Naumov&Naumov,Naumov}, namely
\beqa\label{eq:covariant-gaussian-profile}
F_{P}(K) &=& \mathcal{N} e^{-\scalebox{1.1}{$\frac{P{\cdot}K}{2D^2}$}}
\quad \text{with} \quad P^T \,\equiv\, \sqrt{\mathbf{P}^2 + M^2}\,,
\eeqa
where $D > 0$ is the momentum variance and
\beqa\label{eq:normalisation-mi}
\mathcal{N} &\equiv& \frac{2\pi}{D\,\sqrt{K_1\hspace{-1.0mm}\left(\frac{M^2}{D^2}\right)}}\,,
\eeqa
where $K_1(z)$ stands for the modified Bessel function of the second kind, we obtain
\beqa\label{eq:gwp-m}
\varphi_{X,P}(x) &=& \frac{\mathcal{N}M^2}{(2\pi)^2}\,
\frac{K_1\hspace{-0.8mm}\left(\frac{M^2}{D^2}\big(\frac{1}{4}
+i\,\frac{D^2}{M^2}\,P{\cdot}(x-X) - \frac{D^4}{M^2}\,(x-X)^2\big)^\frac{1}{2}\right)}{\frac{M^2}{D^2}
\left(\frac{1}{4}+i\,\frac{D^2}{M^2}\,P{\cdot}(x-X) - \frac{D^4}{M^2}\,(x-X)^2\right)^\frac{1}{2}}\,.
\eeqa
It follows from $\varphi_{X,P}(x) \propto |\Delta\mathbf{x}|^{-3}$ for $|\mathbf{x}| \gg |\mathbf{X}|$ and
$(\Box + M^2)\varphi_{X,P}(x) = 0$ that the creation
operator $\hat{a}^\dagger(\varphi_{X,P})$ is time-independent in the linear quantum field
theory~\cite{Lehmann&Symanzik&Zimmermann}.

\subsubsection{Wave-packet position in Minkowski spacetime}
\label{sec:gwp-p}

The trajectory of a freely-moving particle in Minkowski spacetime is a straight line. The same result holds
for the trajectory of the wave packet $\varphi_{X,P}(x)$ (cf.~\cite{Newton&Wigner}):
\beqa\label{eq:gwp-m-position}\hspace{-3.5mm}
\langle \mathbf{x}(t) \rangle &\equiv& - i{\int_t}d^3\mathbf{x}\;\mathbf{x}\,
\big(\varphi_{X,P}(x)\partial_t \overline{\varphi}_{X,P}(x)
- \overline{\varphi}_{X,P}(x)\partial_t\varphi_{X,P}(x)\big)
\;=\; \mathbf{X} + \langle \mathbf{V} \rangle\,(t - T)\,,
\eeqa
where
\beqa
\langle \mathbf{V} \rangle &\equiv& \frac{1}{2}{\int}\frac{d^3\mathbf{K}}{(2\pi)^3}\,
\frac{|F_{\mathbf{P}}(\mathbf{K})|^2}{\sqrt{\mathbf{K}^{2} + M^2}}\,
\frac{\mathbf{K}}{\sqrt{\mathbf{K}^{2} + M^2}}
\;\xrightarrow[M/D \,\rightarrow\, \infty]{}\; \frac{\mathbf{P}}{\sqrt{\mathbf{P}^2 + M^2}}\,.
\eeqa
Thus, $\varphi_{X,P}(x)$ propagates like a classical (point-like) particle
of the same mass if $M \gg D$.

\subsubsection{Wave-packet momentum in Minkowski spacetime}
\label{sec:gwp-m}

Making use of $[\hat{\Phi}(x),\hat{\Phi}(x')] = i\Delta(x-x')\hat{1}$, where
$\Delta(x-x')$ is the commutator function, we find the stress-tensor expectation
value in the single-particle state $|\varphi_{X,P}\rangle$:
\beqa
\langle\hat{\Theta}_{\mu\nu}\rangle &=&
2\partial_{(\mu}\overline{\varphi}_{X,P}\,\partial_{\nu)}\varphi_{X,P}
- \frac{1}{3}\,\partial_\mu\partial_\nu|\varphi_{X,P}|^2 -
\frac{1}{3}\,\eta_{\mu\nu}\big(|\partial\varphi_{X,P}|^2 - M^2|\varphi_{X,P}|^2\big)\,,
\eeqa
where we have omitted the vacuum stress tensor, as this does not
depend on the wave packet.
The energy and momentum, which are ascribed to the wave packet, are given by
\bsubeqs\label{eq:gwp-m-momentum}
\beqa
\langle p_t(t)\rangle &\equiv& {\int_t}d^3\mathbf{x}\,\langle\hat{\Theta}_{t}^t(x)\rangle
\;=\; \left({K_2}\hspace{-0.8mm}\left(\scalebox{1.2}{$\frac{M^2}{D^2}$}\right)/
{K_1}\hspace{-0.8mm}\left(\scalebox{1.2}{$\frac{M^2}{D^2}$}\right)\right) P_t
\;\xrightarrow[M/D \,\rightarrow\, \infty]{}\; P_t\,,
\\[1mm]\label{eq:3-momentum}
\langle p_i(t)\rangle &\equiv& {\int_t}d^3\mathbf{x}\,\langle\hat{\Theta}_{i}^t(x)\rangle
\;=\; \left({K_2}\hspace{-0.8mm}\left(\scalebox{1.2}{$\frac{M^2}{D^2}$}\right)/
{K_1}\hspace{-0.8mm}\left(\scalebox{1.2}{$\frac{M^2}{D^2}$}\right)\right) P_i
\;\xrightarrow[M/D \,\rightarrow\, \infty]{}\; P_i\,.
\eeqa
\esubeqs
The packet $\varphi_{X,P}(x)$ is thus characterised by the four-momentum
like a classical (point-like) particle of the same mass and three-momentum if $M \gg D$.

\subsection{Covariant wave packet in de-Sitter spacetime}

The phase factor~\eqref{eq:covariant-phase-factor} suggests that a
single-particle wave packet in curved spacetime must be covariant.
Its Fourier-transform representation should then have the following structure:
\beqa\label{eq:lwp-cs}
\varphi_{X,P}(x) &=& {\int}\frac{d^4K}{(2\pi)^3}\,\theta(K^T)\,
\delta\big(K_AK^A-M^2\big)
F_{P}(K)\,\phi_{X,K}(x)\,,
\eeqa
where the (upper) index $A$ refers to the tangent frame at $X$ with the vierbein
$e_A^{M}(X)$, and
\beqa\label{eq:lpw-cs}
\phi_{X,K}(x) &\xrightarrow[]{\text{$x$ close to $X$}}& e^{iK{\cdot}\sigma}\,,
\eeqa
where $\sigma$ is a shorthand notation for the geodetic distance $\sigma(x,X)$, so that
\beqa
K{\cdot}\sigma &\equiv& K_{A}\sigma^{A} \;=\; e_A^{M}(X)e_{N}^A(X) K_M\sigma^N
\;=\; K_M\sigma^M\,,
\eeqa
where
\beqa
\sigma^M &\equiv& \nabla^M\sigma(x,X) \;=\; g^{MN}(X)\,\partial_N\sigma(x,X)
\eeqa
is a vector of length equal to
the distance along the geodesic between $x$ and $X$, tangent to it at $X$,
and oriented in the direction from $x$ to $X$~\cite{DeWitt}. Note, in general,
$\phi_{X,K}(x)$ is a function of
dimensionless combinations of the curvature tensor at $X$ with $K_{M}$ and
$\sigma_{M}$.

The covariant wave packet can depend only
on $K{\cdot}\sigma$ and $\sigma$, which is due to the de-Sitter-spacetime symmetries.
Without loss of generality, we look for a solution of~\eqref{eq:sfe}~in~the~form
\beqa\label{eq:phi-ds}
\phi_{X,K}(x) &=&
\frac{1}{4\pi i}\,\frac{\sqrt{2H^2\sigma}}{{\sinh}\sqrt{2H^2\sigma}}\,
\frac{ \phi(K{\cdot}\sigma,R\sigma)}{\sqrt{K{\cdot}\sigma^2 - 2K^2\sigma}}\,,
\eeqa
where the Ricci scalar $R = - 12H^2$. Substituting~\eqref{eq:phi-ds} in the field
equation~\eqref{eq:sfe} and using
\bsubeqs
\beqa
\sigma_{;\mu}^{;\mu} &=& 1 + 3\sqrt{2H^2\sigma}\,{\coth}\sqrt{2H^2\sigma}\,,
\\[3.5mm]
\sigma^\mu (K{\cdot}\sigma)_{;\mu} &=& K{\cdot}\sigma\,,
\\[2.5mm]
(K{\cdot}\sigma)_{;\mu}^{;\mu} &=& 3H^2\frac{K{\cdot}\sigma}{\sqrt{2H^2\sigma}}
\bigg[{\coth}\sqrt{2H^2\sigma} - \frac{\sqrt{2H^2\sigma}}{{\sinh^2}\sqrt{2H^2\sigma}}\bigg]\,,
\\[1mm]
(K{\cdot}\sigma)^{;\mu}(K{\cdot}\sigma)_{;\mu} &=& \frac{2H^2\sigma}{{\sinh^2}\sqrt{2H^2\sigma}}\,
K^2 + \frac{K{\cdot}\sigma^2}{2\sigma}\bigg[1-\frac{2H^2\sigma}{{\sinh^2}\sqrt{2H^2\sigma}}\bigg]\,,
\eeqa
\esubeqs
we obtain on the mass shell, $K^2 = M^2$, that
\beqa\label{eq:chi-eq}
\bigg(\frac{\partial^2}{\partial \eta^2}-\frac{\partial^2}{\partial \zeta^2}
+ \frac{\gamma(1-\gamma)}{{\sinh^2}\eta}\bigg) \phi(\eta,\zeta) &=& 0 \quad \text{with} \quad
\gamma \;\equiv\; \frac{1}{2}\big(1-i\sqrt{4\nu^2-1}\big)\,,
\eeqa
where we have introduced new variables, namely
\bsubeqs
\beqa\label{eq:eta}
\eta &\equiv& \ln\tanh\bigg[\frac{\sqrt{2H^2\sigma}}{2}\bigg]\,,
\\[1mm]
\zeta &\equiv& \ln\bigg[\frac{\sqrt{K{\cdot}\sigma^2 - 2K^2\sigma}-K{\cdot}\sigma}{\sqrt{2K^2\sigma}}\bigg]\,,
\eeqa
\esubeqs
such those $\eta \in (-\infty,0)$ and $\zeta \in [0,+\infty)$ if $\sigma > 0$ and
$-K{\cdot}\sigma \geq \sqrt{2K^2\sigma}$ are fulfilled.

\subsubsection{Locally plane-wave solutions in de-Sitter spacetime}

The equation~\eqref{eq:chi-eq} has infinitely many solutions. One of them reads
\beqa\label{eq:chi-sol}
 \phi_\nu(\eta,\zeta) &\equiv& {\int\limits_{-\infty}^{+\infty}}dp\, e^{(ip-1)\zeta}
\big( \phi_{\nu,\,ip-1}(\eta) -  \phi_{\nu,\,1-ip}(\eta)\big)\,,
\eeqa
where by definition
\beqa
 \phi_{\nu,\,ip-1}(\eta) &\equiv&  e^{(ip-1)(\eta + \ln i\nu)}\,
\frac{\Gamma[\scalebox{0.90}{$2 - \gamma- ip$}]\Gamma[\scalebox{0.90}{$1+\gamma - ip$}]}
{\Gamma[\scalebox{0.90}{$1-ip$}]}\,
{}_2F_1\bigg[\gamma,1-\gamma;\scalebox{0.90}{$2-ip$};\frac{1}{1-e^{2\eta}}\bigg].
\eeqa

The coefficient to depend only on $p$ and $\nu$ has been chosen from the following
argument. In the observable Universe, $M \ggg H_0$ holds for the massive fields of
the Standard Model of elementary particle physics. It is an empirical fact that collider
physics is well described by the Minkowski plane-wave solutions.
We must, therefore, obtain a plane-wave solution
$e^{i K{\cdot}\sigma}$ for $\phi_{X,K}(x)$ if $H \lll M$ and $H^2|\sigma| \ll 1$ are
satisfied. Specifically, if $\nu \equiv M/H \rightarrow \infty$,~then~we find from the definition
of the hypergeometric function and 8.328.2 in~\cite{Gradshteyn&Ryzhik} that
\beqa
{}_2F_1\big(\gamma,1-\gamma;c;-|x|\big) &\xrightarrow[\nu \,\rightarrow\, \infty]{}&
\Gamma[c]\big(\nu\sqrt{|x|}\big)^{1-c}J_{c-1}\big(2\nu\sqrt{|x|}\big)\,,
\eeqa
where $J_{c-1}(z)$ is the Bessel function of the first kind.
This result agrees with 10.16.10 in~\cite{Olver&Lozier&Boisvert&Clark}.
Employing 9.131 in~\cite{Gradshteyn&Ryzhik}, we find that
\beqa\nonumber
{}_2F_1\bigg[\gamma,1-\gamma;\scalebox{0.90}{$2-ip$};\frac{1}{1-e^{2\eta}}\bigg]
&\xrightarrow[\nu \,\rightarrow\, \infty]{}&
\frac{\Gamma[\scalebox{0.90}{$1- ip$}]}{\Gamma[\scalebox{0.90}{$2-\gamma-ip$}]
\Gamma[\scalebox{0.90}{$1+\gamma-ip$}]}\,\frac{\pi i(ip-1)}{\sinh(\pi p)}\,e^{(1-ip)\eta}\nu^{1-ip}
\\[1mm]
&&\hspace{-47mm}
\,{\times}\bigg[\big(1-e^{2\eta}\big)^{\frac{ip-1}{2}}J_{ip-1}\Big(\frac{2\nu e^\eta}{\sqrt{1-e^{2\eta}}}\Big)
+e^{\pi p}\big(1-e^{2\eta}\big)^{\frac{1-ip}{2}}J_{1-ip}\Big(\frac{2\nu e^\eta}{\sqrt{1-e^{2\eta}}}\Big)\bigg].
\eeqa
If we consider $H^2|\sigma| \ll 1$, then $\eta$ approaches $-\infty$, i.e. we are
allowed to set $1-e^{2\eta}$ to unity in this limit, whereas $2\nu e^\eta$ turns into
$\sqrt{2M^2\sigma}$. Having used 10.4.8 in~\cite{Olver&Lozier&Boisvert&Clark}, we obtain
\beqa
 \phi_\nu(\eta,\zeta) &\xrightarrow[H \,\lll\, M]{H^2|\sigma| \,\ll\, 1}&
-2\pi{\int\limits_{-\infty}^{+\infty}}dp\,(ip-1)e^{(ip-1)\zeta}e^{\frac{\pi p}{2}}H_{ip-1}^{(2)}
\big(\sqrt{2M^2\sigma}\big)\,,
\eeqa
where $H_{ip-1}^{(2)}(z)$ is the Hankel function of the second kind. With the help of~8.421.2
in~\cite{Gradshteyn&Ryzhik}, we have that
\beqa
 \phi_\nu(\eta,\zeta) &\xrightarrow[H \,\lll\, M]{H^2|\sigma| \,\ll\, 1}&
4\pi i\sqrt{2M^2\sigma}\,\sinh \zeta\,e^{-i\sqrt{2M^2\sigma}\,\cosh \zeta}\,.
\eeqa
Substituting this result into~\eqref{eq:phi-ds}, we find
\beqa
\phi_{X,K}(x)
&\xrightarrow[H \,\lll\, M]{H^2|\sigma| \,\ll\, 1}&
e^{i K{\cdot}\sigma}\,,
\eeqa
as required.

The integral over $p$ in \eqref{eq:chi-sol} can actually be exactly evaluated. Specifically,
we obtain from 15.6.7 in~\cite{Olver&Lozier&Boisvert&Clark}
and~6.422.12 in~\cite{Gradshteyn&Ryzhik} that
\beqa
 \phi_\nu(\eta,\zeta) &=& \lim_{w \,\rightarrow\, 0}\,\partial_w\Big(
\Phi_\nu\big(\eta,w+\zeta\big) - \Phi_\nu\big(\eta,w-\zeta\big)\Big)\,,
\eeqa
where by definition
\beqa
\Phi_\nu(\eta,\zeta) &\equiv& -2\sqrt{2\pi\nu}\,\sqrt{-e^\zeta\sinh \eta}\,e^{-\frac{\pi i}{4}}
e^{-i\nu e^\zeta{\cosh}\,\eta}K_{\gamma-\frac{1}{2}}\big(i\nu e^\zeta{\sinh}\,\eta\big)\,.
\eeqa
Considering now the limit $\nu \rightarrow \infty$ and then $H^2|\sigma| \ll 1$, we obtain
$\phi_{X,K}(x) \rightarrow e^{i K{\cdot}\sigma}$,
where~we have made use of the uniform expansion of $K_{i\nu}(\nu z)$ for $\nu \rightarrow \infty$ found
in~\cite{Balogh}.

The $\phi_{\nu,\,1-ip}(\eta)$-dependent part of the integrand in~\eqref{eq:chi-sol}
vanishes in the limit $\nu \rightarrow 0$. We therefore consider in the case $\nu \equiv M/H = 0$ that
\beqa
\phi_0(\eta,\zeta) &\equiv& 2\lim_{\nu \,\rightarrow\, 0}
{\int\limits_{-\infty}^{+\infty}}dp\, e^{(ip-1)\zeta}\phi_{\nu,\,ip-1}(\eta) \;=\;
2\lim_{\nu \,\rightarrow\, 0}{\int\limits_{-\infty}^{+\infty}}dp\,e^{(ip-1)(\zeta + \eta + \ln i\nu)}\,
\Gamma[\scalebox{0.90}{$2-ip$}]\,.
\eeqa
Taking into account~3.328 in~\cite{Gradshteyn&Ryzhik}, we obtain in the massless ($M = 0$) case that
\beqa\label{eq:massless-solution}
\phi_{X,K}(x) &=& \frac{2}{{\cosh}\sqrt{2H^2\sigma} + 1}\,
\exp\bigg(iK{\cdot}\sigma\,\frac{\tanh\frac{1}{2}\sqrt{2H^2\sigma}}{\frac{1}{2}\sqrt{2H^2\sigma}}\bigg).
\eeqa
Note, $\phi_{X,K}(x)$ turns into the standard plane-wave solution $e^{i K{\cdot}\sigma}$ as
in Minkowski spacetime if $H^2|\sigma| \ll 1$ holds. This simple
form of $\phi_{X,K}(x)$ explains our minimal Ansatz for the on-shell condition we have assumed 
on physical grounds
in~\eqref{eq:lwp-cs}.

\subsubsection{In-in and in-out propagators in de-Sitter spacetime}
\label{sec:ii-io}

To clarify a non-trivial structure of the $\eta$-dependent integrand in~\eqref{eq:chi-sol}, we need
to compute the Wightman function that might be associated with this solution.
In Minkowski spacetime, $H = 0$, the two-point function can be found as follows:
\beqa
W(x,X)\big|_{H = 0} &=& {\int}\frac{d^4K}{(2\pi)^3}\,\theta(K^T)\,\delta(K^2-M^2)\,e^{i K{\cdot}\sigma}\,.
\eeqa
In the de-Sitter universe, the correlation function may be defined via the same
formula with $e^{i K{\cdot}\sigma}$ replaced by $\phi_{X,K}(x)$, where
$K$ belongs to the cotangent space at $X$:
\beqa\nonumber\label{eq:w-in-in}
W(x,X) &=& \frac{1}{4i(2\pi)^4\nu}\,\frac{1}{{\sinh}\sqrt{2H^2\sigma}}
{\int\limits_{-\infty}^{+\infty}}dp\,\big(\phi_{\nu,\,ip-1}(\eta) - \phi_{\nu,\,1-ip}(\eta)\big)
{\int}\frac{d^3\mathbf{K}}{\sqrt{\mathbf{K}^2 + M^2}}\,\frac{e^{(ip-1)\zeta}}{\sinh \zeta}
\\[1mm]\nonumber
&=& \frac{H^2}{4i(2\pi)^3}\frac{\sqrt{\gamma(1{-}\gamma)}}{{\sinh}\sqrt{2H^2\sigma}}
{\int\limits_{-\infty}^{+\infty}}dp\,\big(\phi_{\nu,\,ip-1}(\eta) - \phi_{\nu,\,1-ip}(\eta)\big){\int\limits_{0}^\infty}dq\,
\frac{e^{ipq} + e^{(ip-2)q}}{1-ip}
\\[1mm]\label{eq:wf-ds}
&=& -\frac{H^2}{8\pi}\,\frac{\gamma(1{-}\gamma)}{{\sin}\pi\gamma}\,
\frac{1}{{\cosh}\sqrt{2H^2\sigma} -1}\,
{}_2F_1\bigg[\gamma,1-\gamma;2;\frac{1+{\cosh}\sqrt{2H^2\sigma}}{2}\bigg].
\eeqa
Note, in the second line, one can replace $+p$ by $-p-2i$ in the $e^{(ip-2)q}$-dependent part of the
integrand to get the delta function $\delta(p)$ from the integral over $q$, by
taking into account that residues at $\pm\mu -3i/2$ and $\mp\mu -i/2$ cancel
each other and $\phi_{\nu,\,\pm ip \mp1}(\eta)$ vanishes exponentially in the limit
$\text{Re}\,p \rightarrow \pm\infty$.

Therefore, $\phi_\nu(\eta,\zeta)$ might be related to the Wightman function of the
Chernikov-Tagirov aka Bunch-Davies state~\cite{Chernikov&Tagirov,Bunch&Davies}.
If we assume that $\text{Im}({\cosh}\sqrt{2H^2\sigma}) < 0$, then the correlation
function~\eqref{eq:w-in-in} turns into the in-in propagator. It was, however, argued
in~\cite{Polyakov} that the in-out propagator should be considered in non-linear
quantum field models in de-Sitter spacetime. This type of the Feynman propagator
is associated with $\phi_\nu(+\eta,\zeta) + e^{-\pi i \gamma}\phi_\nu(-\eta,\zeta)$,
which reduces to $\phi_\nu(+\eta,\zeta)$ if $\nu \rightarrow \infty$. Thus, the solution
we were looking for is not unique.

In the massless case, $M = 0$, the integration over the Fourier parameters $K$ in the formula
of the Wightman function gives the result~\eqref{eq:wf-ds} if taken in the limit $\nu \rightarrow 0$.
The discontinuity $\phi_0(\eta,\zeta) \neq \phi_{\nu \rightarrow 0}(\eta,\zeta)$ is also
present in the integration over the Fourier parameters. Namely, the integral over $\mathbf{K}$
in the first line of~\eqref{eq:w-in-in} is alone proportional to the delta function $\delta(p)$~if $\nu = 0$,
while not if $\nu > 0$.

\subsubsection{Gaussian wave packet in de-Sitter spacetime}

According to our suggestion, the Gaussian wave packet in de-Sitter spacetime reads
\beqa\label{eq:gwp-ds}
\varphi_{X,P}(x) &=& \mathcal{N}{\int}\frac{d^4K}{(2\pi)^3}\,\theta\big(K^T\big)\,
\delta\big(K^2-M^2\big)\,
e^{-\scalebox{1.1}{$\frac{P{\cdot}K}{2D^2}$}}\,\phi_{X,K}(x)\,,
\eeqa
where $D$ is the momentum variance and $\mathcal{N}$ needs to be determined from
the normalisation condition
\beqa
- i{\int_\Sigma}d\Sigma^\mu(x)\,
\big(\varphi_{X,P}(x)\nabla_\mu\overline{\varphi}_{X,P}(x) -
\overline{\varphi}_{X,P}(x)\nabla_\mu\varphi_{X,P}(x)\big) &=& 1\,,
\eeqa
where $\Sigma$ is a Cauchy surface. Since the wave packet $\varphi_{X,P}(x)$
vanishes as $|\Delta\mathbf{x}|^{-3}$ for large~$|\Delta\mathbf{x}|$ and is a
solution of the scalar-field equation~\eqref{eq:sfe}, the normalisation factor does not
depend~on the Cauchy surface. Therefore, it generically holds
$\mathcal{N} = \mathcal{N}(M,D,H)$ (see fig.~\ref{fig:normalisation-position}, left).

Plugging $\phi_{X,K}(x)$ found above into~\eqref{eq:gwp-ds} and assuming that $M/D > 0$, we obtain
\beqa\label{eq:gwf-ds}
\varphi_{X,P}(x) &=& \frac{iM^2\mathcal{N}}{4\nu(2\pi)^3}
{\int\limits_{-\infty}^{+\infty}}dw\sinh w\,e^{-\frac{M^2}{2D^2}\,{\cosh}\,w}\,
\frac{\Phi_\nu(\eta,w+\upsilon) - \Phi_\nu(\eta,w-\upsilon)}{\csch \eta\,\sinh \upsilon}\,,
\eeqa
where $\eta$ has been defined in~\eqref{eq:eta} and by definition
\beqa
\upsilon &\equiv& \ln\bigg[\frac{\sqrt{P{\cdot}\sigma^2 - 2M^2\sigma}-P{\cdot}\sigma}
{\sqrt{2M^2\sigma}}\bigg]\,.
\eeqa
The integral over $w$ in~\eqref{eq:gwf-ds} seems not to be generically tractable. Still, it can be ``simplified"
with the help of~8.432.1 in~\cite{Gradshteyn&Ryzhik} and the first formula on
p.~86 in~\cite{Magnus&Oberhettinger&Soni}.

In Secs.~\ref{sec:gwp-p} and~\ref{sec:gwp-m}, we have learned that the Gaussian wave packet in
Minkowski spacetime kinematically behaves as a classical point-like particle if its mass $M$ is much
larger than its momentum variance $D$. Considering $M \gg D$ in~\eqref{eq:gwf-ds}, we observe
that the integrand is extremely suppressed for $|w| \gtrsim 1$. Therefore, if we multiply
that integrand by $\exp(-\frac{1}{2}w)$, then \eqref{eq:gwf-ds} remains essentially unchanged in
the case $M \gg D$. However, this modified integral can be exactly evaluated by
using~6.653.2 in~\cite{Gradshteyn&Ryzhik}.
Specifically, we get
\beqa\label{eq:mgwf-ds}
\tilde{\varphi}_{X,P}(x) &\equiv& \frac{iM^2\mathcal{N}}{4\nu(2\pi)^3}
{\int\limits_{-\infty}^{+\infty}}dw\sinh w\,e^{-\frac{M^2}{2D^2}\,{\cosh}\,w -\frac{1}{2}w}\,
\frac{\Phi_\nu(\eta,w+\upsilon) - \Phi_\nu(\eta,w-\upsilon)}{\csch \eta\,\sinh \upsilon}
\\[1mm]\nonumber
&=& \frac{2H^2\mathcal{N}c_+ c_-}{(2\pi)^{\frac{5}{2}}(c_+ - c_-)}\Big(
\sqrt{c_{+}}\,\partial_{b_+}\big(K_{i\mu}(\chi_+^+)K_{i\mu}(\chi_+^-)\big)-
\sqrt{c_{-}}\,\partial_{b_-}\big(K_{i\mu}(\chi_-^+)K_{i\mu}(\chi_-^-)\big)
\Big)\,,
\eeqa
where by definition
\bsubeqs
\beqa
\chi_{\pm}^+ &\equiv& \frac{1}{2}\big(a_\pm - b_\pm\big)^\frac{1}{2}\Big(
\big(a_\pm + b_\pm + 2c_\pm\big)^\frac{1}{2} +
\big(a_\pm + b_\pm - 2c_\pm\big)^\frac{1}{2}
\Big)\,,
\\[1mm]
\chi_{\pm}^- &\equiv& \frac{1}{2}\big(a_\pm - b_\pm\big)^\frac{1}{2}\Big(
\big(a_\pm + b_\pm + 2c_\pm\big)^\frac{1}{2} -
\big(a_\pm + b_\pm - 2c_\pm\big)^\frac{1}{2}
\Big)\,,
\eeqa
\esubeqs
and
\bsubeqs
\beqa
a_\pm &\equiv& \frac{M^2}{2D^2} + b_\pm\,,
\\[1mm]
b_\pm &\equiv& i\nu e^{\pm v}\cosh\eta\,,
\\[1mm]
c_\pm &\equiv& i\nu e^{\pm v}\sinh\eta\,.
\eeqa
\esubeqs
Numerical computations with $\varphi_{X,P}(x)$ and $\tilde{\varphi}_{X,P}(x)$ give us the same
results within numerical error bars. However, it is worth mentioning at this point that the
integrand in~\eqref{eq:gwf-ds} is highly oscillatory. Presumably, this circumstance makes it
non-trivial to do numerics with $\varphi_{X,P}(x)$ if used its integral form.

\subsubsection{Wave-packet position in de-Sitter spacetime}
\begin{figure}
\centering
\includegraphics[scale=0.9]{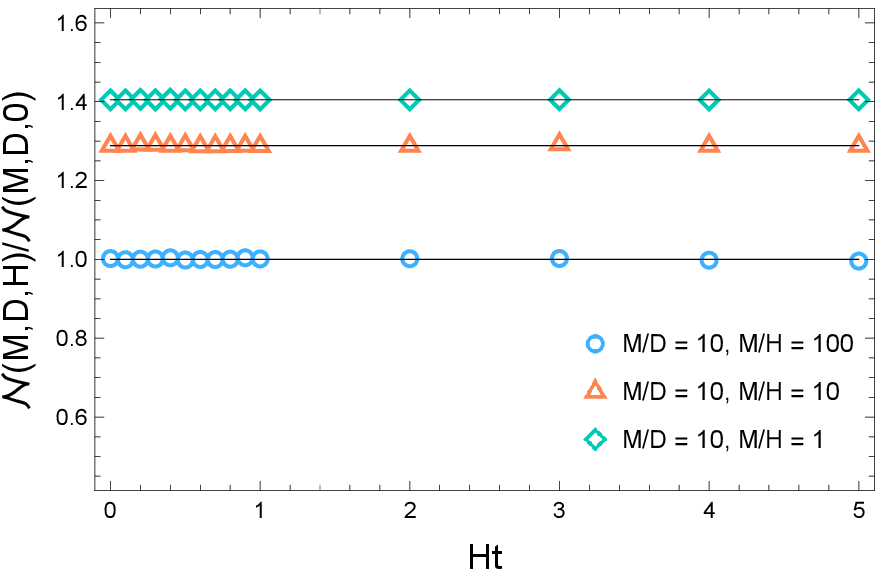}\hspace{2mm}
\includegraphics[scale=0.9]{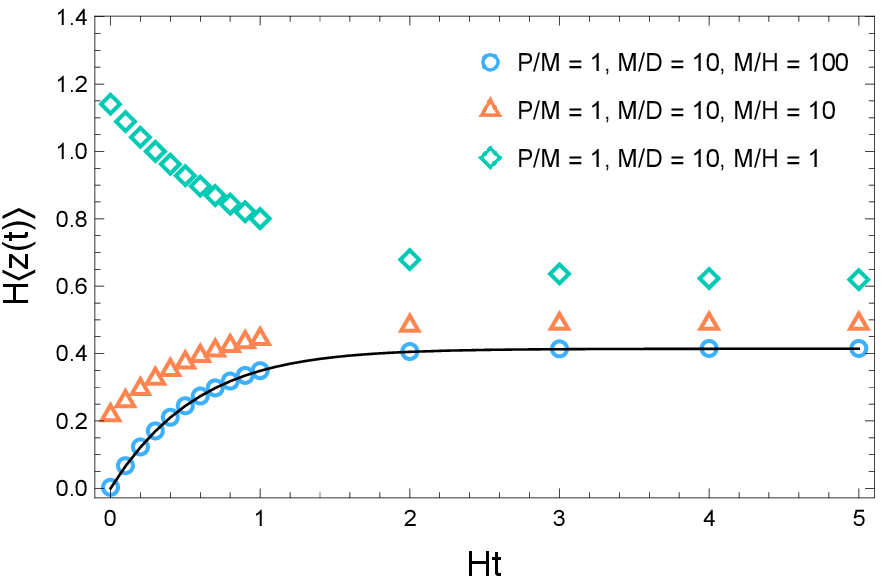}\vspace{2mm}
\caption{Left: Numerical calculation of the normalisation factor $\mathcal{N}(M,D,H)$.
This plot shows the ratio $\mathcal{N}(M,D,H)/\mathcal{N}(M,D,0)$, where
$\mathcal{N}(M,D,0)$ is the Minkowski-spacetime normalisation factor
(see equation~\eqref{eq:normalisation-mi}).
Right: Numerical calculation of $\langle z(t) \rangle$, where
the black solid curve~corresponds to the classical trajectory $z(t)$
(see equation~\eqref{eq:classical-trajectory}).}\label{fig:normalisation-position}
\end{figure}

Without loss of generality, we intend to consider a free motion with the initial conditions
$X = 0$ and $P = (\sqrt{M^2 + P^2},0,0,P)$ (in the tangent frame at $X$). The position
of a classical particle of the mass $M$ in this case reads
\beqa\label{eq:classical-trajectory}
z(t) &=& \frac{1}{PH}\Big(\sqrt{M^2 + P^2}-\sqrt{M^2 + e^{-2Ht}P^2}\Big)\,,
\eeqa
where $x(t) = y(t) = 0$ due to the spatial-translation symmetry of the flat de-Sitter universe.

In analogy to the Minkowski case, the wave-packet position should follow from
\beqa\nonumber\label{eq:position-ev}
\langle z(t) \rangle &\equiv& - i{\int_\Sigma}d\Sigma^\mu(x)\,z\,
\big(\varphi_{X,P}(x)\nabla_\mu\overline{\varphi}_{X,P}(x) -
\overline{\varphi}_{X,P}(x)\nabla_\mu\varphi_{X,P}(x)\big)
\\[1mm]
&=& -ie^{3Ht}{\int_t}d^3\mathbf{x}\,z\,
\big(\varphi_{X,P}(x)\partial_t\overline{\varphi}_{X,P}(x) -
\overline{\varphi}_{X,P}(x)\partial_t\varphi_{X,P}(x)\big)\,,
\eeqa
whereas $\langle x(t) \rangle = \langle y(t) \rangle = 0$ due to the
invariance of $\varphi_{X,P}(x)$ under rotations around $z$-axis.
It should be mentioned that $\varphi_{X,P}(x)$ is spherically symmetric if $P = 0$.
In this special case, we immediately obtain $z(t) = \langle z(t) \rangle = 0$.

In general, we
numerically find that $\langle z(t) \rangle$ matches the classical
trajectory if $M \gg D \gg H$ (see fig.~\ref{fig:normalisation-position}, right).
If the value of the Hubble parameter $H$ is comparable with either~the momentum
variance $D$ or the scalar-field mass $M$, then the wave-packet trajectory
differs~from the classical geodesic~\eqref{eq:classical-trajectory}. Specifically,
the initial wave-packet position deviates from~$X = 0$ if $H$
is non-negligible with respect to $D$. Besides, the wave-packet propagation rate
decreases with increasing $H$ and is negative if $H \sim M$.

\subsubsection{Wave-packet momentum in de-Sitter spacetime}
\begin{figure}
\centering
\includegraphics[scale=0.9]{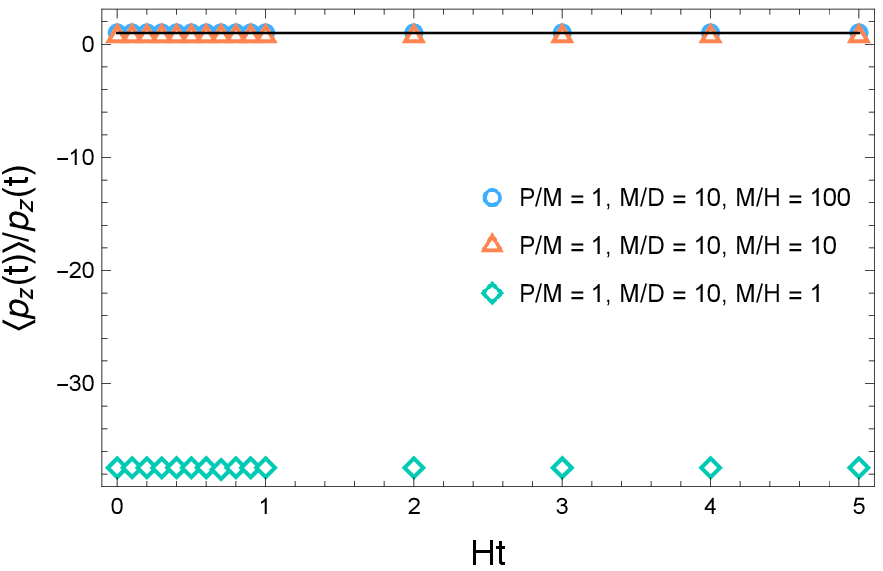}\hspace{2mm}
\includegraphics[scale=0.9]{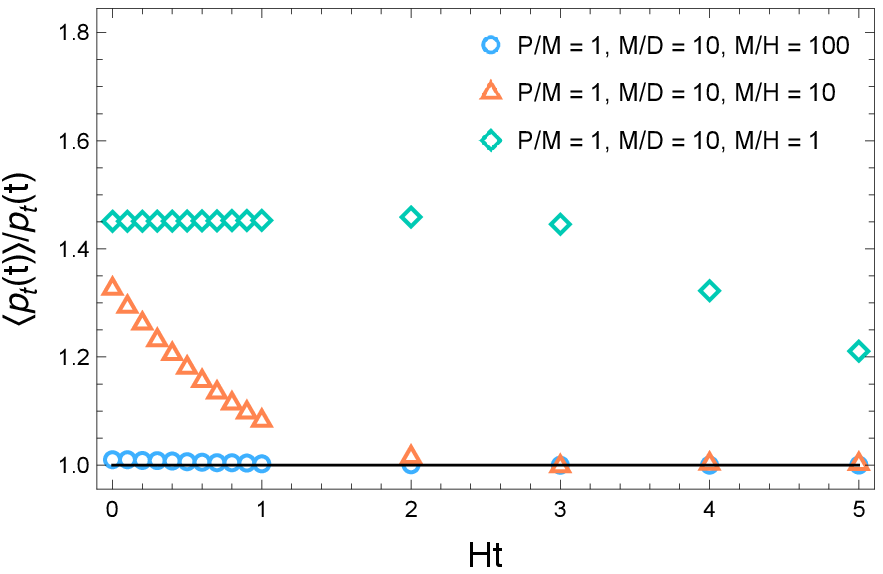}\vspace{2mm}
\caption{Left: Numerical calculation of $\langle p_z(t)\rangle$.
The solid straight line corresponds to the classical result (see equation~\eqref{eq:cl-momentum}).
Right:  Numerical calculation of $\langle p_t(t)\rangle$. This plot
shows $\langle p_t(t)\rangle/p_t(t)$, where $p_t(t)$ corresponds to the 
classical energy (equation~\eqref{eq:cl-energy}) with $P$ replaced by
the wave-packet momentum $\langle p_z(t)\rangle$. In the case of $M/H = 1$, the ratio
$\langle p_t(t)\rangle/p_t(t)$ appears to be oscillating around $1.165$ at
$Ht \in \{6,7,\dots,19,20\}$ with the amplitude $0.018$.}\label{fig:energy-momentum}
\end{figure}

The four-momentum of the classical particle can be directly found to read
\bsubeqs\label{eq:cl-four-momentum}
\beqa\label{eq:cl-energy}
p^t(t) &=& \sqrt{M^2 + e^{-2Ht}P^2}\,,
\\[1mm]\label{eq:cl-momentum}
p^z(t) &=& e^{-2Ht}P\,,
\eeqa
\esubeqs
where $p^x(t) = p^y(t) = 0$ due to the initial conditions considered and the spatial-translation
symmetry of flat de-Sitter spacetime.

Making use of the commutator function in de-Sitter spacetime, one
finds that the stress-tensor expectation value in the single-particle state $|\varphi_{X,P}\rangle$ is
\beqa\nonumber
\langle\hat{\Theta}_{\mu\nu}\rangle &=&
\nabla_{\mu}\overline{\varphi}_{X,P}\nabla_{\nu}\varphi_{X,P}+
\nabla_{\nu}\overline{\varphi}_{X,P}\nabla_{\mu}\varphi_{X,P}
\\[2mm]&&
-\,\frac{1}{3}\,\nabla_\mu\nabla_\nu|\varphi_{X,P}|^2
-\frac{1}{3}\,g_{\mu\nu}\big(|\nabla\varphi_{X,P}|^2 - (M^2-H^2)|\varphi_{X,P}|^2\big)\,,
\eeqa
where we have omitted the vacuum contribution, as this does not depend on the wave packet.
It is straightforward to show that this energy-momentum tensor is covariantly
conserved, i.e. $\nabla^\mu\langle\hat{\Theta}_{\mu\nu}\rangle = 0$.
Since $\{\partial_i\}$ are three Killing vectors of the de-Sitter universe,
the momentum-conservation law holds, namely
\beqa
\langle p_i(t) \rangle &\equiv& {\int_\Sigma}d\Sigma^\mu(x)\,\langle\hat{\Theta}_{i\mu}(x)\rangle
\;=\; a^3(t){\int_t}d^3\mathbf{x}\,\langle\hat{\Theta}_{i}^t(x)\rangle
\eeqa
does not depend on the Cauchy surface $\Sigma$ (see fig.~\ref{fig:energy-momentum}, left).
Yet, the wave-packet energy,
\beqa
\langle p_t(t) \rangle &\equiv& {\int_\Sigma}d\Sigma^\mu(x)\,\langle\hat{\Theta}_{t\mu}(x)\rangle
\;=\; a^3(t){\int_t}d^3\mathbf{x}\,\langle\hat{\Theta}_{t}^t(x)\rangle
\eeqa
depends generically on cosmic time:
\beqa\label{eq:power}
\frac{d}{dt}\,\langle p_t(t) \rangle
&=& -H\langle p_t(t) \rangle + 2M^2Ha^3(t){\int_t}d^3\mathbf{x}\,|\varphi_{X,P}(t,\mathbf{x})|^2\,,
\eeqa
where we have used $\nabla^\mu\langle\hat{\Theta}_{\mu\nu}\rangle = 0$ and
$\langle\hat{\Theta}_{\mu\nu}\rangle \rightarrow 0$ at spatial infinity (see
fig.~\ref{fig:energy-momentum}, right)

We numerically find that both $\langle p^t(t) \rangle$ and $\langle p^z(t) \rangle$ approach their
classical values~\eqref{eq:cl-four-momentum}~if~the condition $M \gg D \gg H$ is fulfilled, as
shown in fig.~\ref{fig:energy-momentum}. If otherwise, but still keeping $M \gg D$, then the
wave-packet~energy $\langle p^t(t) \rangle$ increases as compared to $p^t(t)$, whereas
$\langle p^z(t) \rangle$ decreases in comparison to $p^z(t)$. These observations seem to hint
that $\langle p^\mu(t)\rangle$ is not a vector, meaning $\langle p^\mu(t)\rangle\langle p_\mu(t)\rangle$
might depend on a coordinate frame. We shall study this issue shortly.

\begin{figure}
\centering
\includegraphics[scale=0.9]{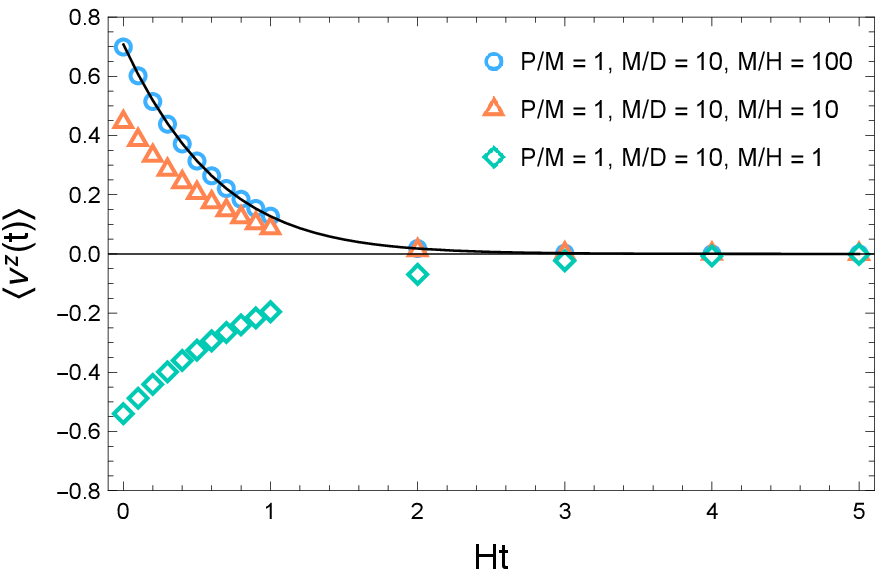}\hspace{2mm}
\includegraphics[scale=0.9]{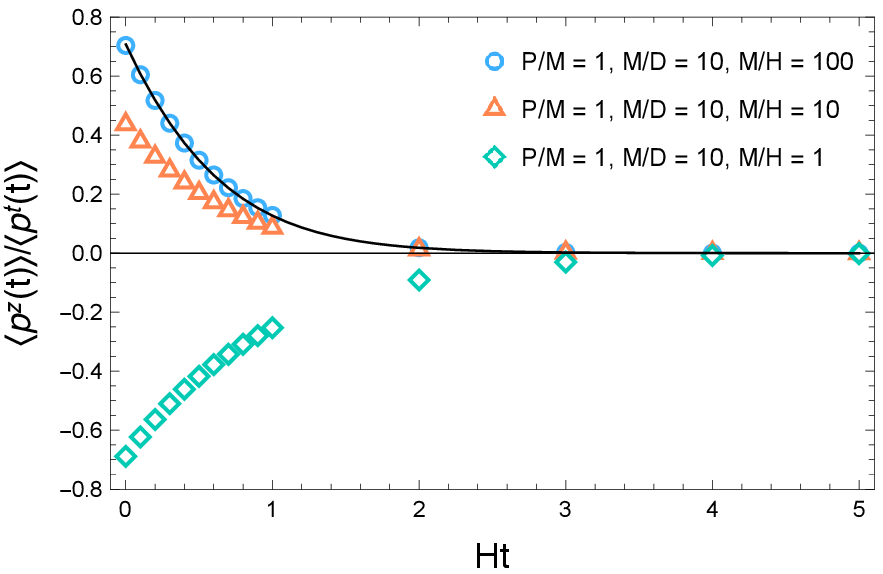}\vspace{2mm}
\caption{Left: Numerical calculation of $\langle \dot{z}(t)\rangle$ which corresponds to the
velocity expectation value~of the wave packet, where the dot stands for the differentiation
with respect to $t$. The classical velocity $\dot{z}(t)$ follows from~\eqref{eq:classical-trajectory}
and is shown by the black solid curve. Right: This plot shows
$\langle p^z(t)\rangle /\langle p^t(t)\rangle$. Point-like particles move with the group velocity,
in particular, $\dot{z}(t) = p^z(t)/p^t(t)$ (the black solid curve). The mismatch between
$\langle \dot{z}(t)\rangle$ and $\langle p^z(t)\rangle /\langle p^t(t)\rangle$ becomes
non-negligible if $M/H \lesssim 10$.}\label{fig:velocities}
\end{figure}

The four-momentum $p^\mu(t)$ of the classical particle is proportional to its
four-velocity~$u^\mu(t)$, namely $p^\mu(t) = M u^\mu(t)$. This formula follows from the
energy-momentum~tensor of a point-like particle whose mass density is given by the
Dirac function~\cite{Landau&Lifshitz}. This relation implies~that
$\dot{z}^z(t) \equiv u^z(t)/u^t(t) = p^z(t)/p^t(t)$, while, in quantum theory, one has
\beqa
\langle \dot{z}(t) \rangle &=& +ie^{Ht}{\int_t}d^3\mathbf{x}\,
\big(\varphi_{X,P}(x)\partial_z\overline{\varphi}_{X,P}(x) -
\overline{\varphi}_{X,P}(x)\partial_z\varphi_{X,P}(x)\big)\,,
\eeqa
where we have taken into account that $\varphi_{X,P}(x)$ is a solution of the field equation~\eqref{eq:sfe},
which tends to zero sufficiently fast at spatial infinity. This integral might now be
expected to~yield the classical result if $M \gg D \gg H \geq 0$ is fulfilled (see fig.~\ref{fig:velocities}).

\subsubsection{Wave-packet position and momentum in co-moving frame}

There is a coordinate frame in de-Sitter spacetime, in which the point-like particle moving along the
geodesic~\eqref{eq:classical-trajectory} with $P \neq 0$ is at rest. This rest coordinate frame can
be readily~found by considering the de-Sitter-hyperboloid embedding into 5D Minkowski spacetime.
Namely, the de-Sitter hyperboloid corresponds to the hypersurface $H^2\eta_{ab}\chi^a\chi^b = -1$,
where $\chi^a$ denote Minkowski coordinates, $\eta_{ab}$ stands for the Minkowski metric and the
indices $a$ and $b$ run from $0$ to $4$. Making use of (A12) in~\cite{Anderson&Mottola}, we obtain
that the diffeomorphism to the rest frame~of~the point-like particle follows from the
Lorentz transformation
\bsubeqs\label{eq:ds-to-rf}
\beqa
\chi^0 &\rightarrow& \tilde{\chi}^0 \;=\; \frac{\sqrt{M^2+P^2}}{M}\,\chi^0 - \frac{P}{M}\,\chi^3\,,
\\[1mm]
\chi^1 &\rightarrow& \tilde{\chi}^1 \;=\; \chi^1\,,
\\[2.5mm]
\chi^2 &\rightarrow& \tilde{\chi}^2 \;=\; \chi^2\,,
\\[1mm]
\chi^3 &\rightarrow& \tilde{\chi}^3 \;=\;  \frac{\sqrt{M^2+P^2}}{M}\,\chi^3 - \frac{P}{M}\,\chi^0\,,
\\[1mm]
\chi^4 &\rightarrow& \tilde{\chi}^4 \;=\; \chi^4\,,
\eeqa
\esubeqs
where $Ht = \ln{H(\chi^0+\chi^4)}$ and $H\mathbf{x} = \boldsymbol{\chi}/(\chi^0 + \chi^4)$.
This coordinate transformation provides $t(\tau) \rightarrow \tau$, $z(\tau) \rightarrow 0$ and
$p(\tau) \rightarrow (M,0,0,0)$, where $\tau$ is the proper time to parametrise points of the
geodesic, $z(\tau)$~and~$p(\tau)$ are, respectively, given in~\eqref{eq:classical-trajectory}
and~\eqref{eq:cl-four-momentum}.

The initial conditions $X = 0$ and $P = (\sqrt{M^2+P^2},0,0,P)$ turn, respectively,
into $\tilde{X} = 0$ and $\tilde{P} = (M,0,0,0)$ in the rest frame in which the covariant
packet is spherically symmetric. In the absence of Lorentz-type deformations, its
position expectation value coincides with the rest-frame origin. Furthermore, the
wave-packet three-momentum also vanishes for the same reason. Hence, the covariant 
wave packet does not propagate in the rest frame,
independent on relative values~of the parameters $M > 0$, $D > 0$ and $H \geq 0$.

The rest-frame position of the packet can be computed from the non-rest-frame
perspective and vice versa. In particular, we have
\beqa\label{eq:tilde-position}
\langle \tilde{x}^\mu(\Sigma) \rangle &=&- i{\int_\Sigma}d\Sigma^\nu(x)\,\tilde{x}^\mu(x)\,
\big(\varphi_{X,P}(x)\nabla_\nu\overline{\varphi}_{X,P}(x) -
\overline{\varphi}_{X,P}(x)\nabla_\nu\varphi_{X,P}(x)\big)\,,
\eeqa
where $\tilde{x}^\mu = \tilde{x}^\mu(x)$ directly follows from~\eqref{eq:ds-to-rf}. It is worth
emphasising that $\langle \tilde{x}^\mu(\Sigma) \rangle$ depends~on the Cauchy surface
$\Sigma$, rather than on $\tilde{\Sigma}$. Still, one might expect that
$\langle \tilde{x}^\mu(\Sigma) \rangle$ approximately coincides with
$\langle \tilde{x}^\mu(\tilde{\Sigma}) \rangle$ if the wave packet manifests itself as
a classical particle.

\begin{figure}
\centering
\includegraphics[scale=0.9]{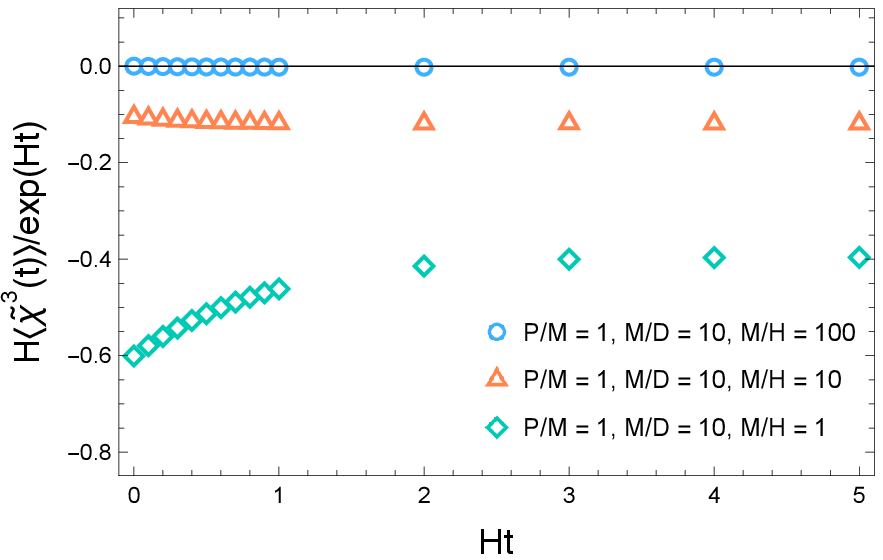}\hspace{2mm}
\includegraphics[scale=0.9]{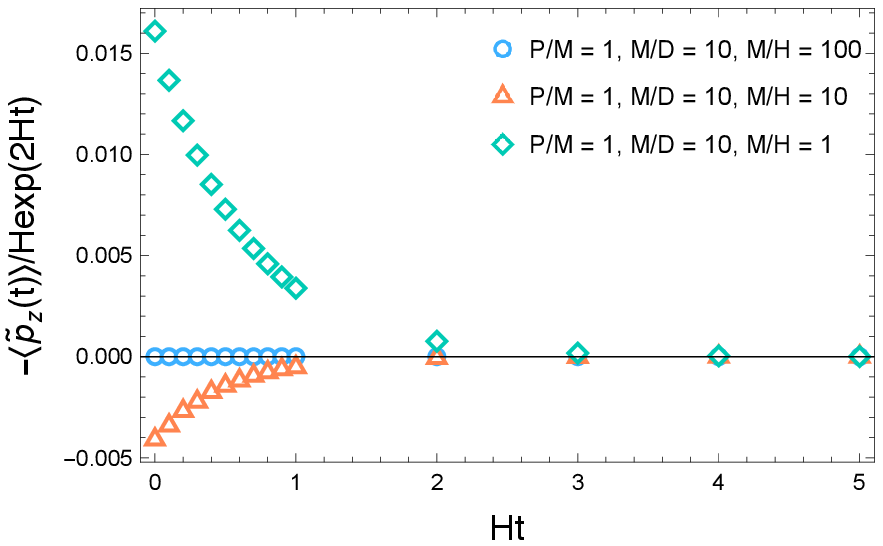}\vspace{2mm}
\caption{Left: Numerical~calculation of $\langle \tilde{\chi}^3(t) \rangle$. In classical theory,
$\tilde{\chi}^3 = \tilde{z}\exp(H\tilde{t})$ implies $\tilde{\chi}^3 = 0$ if $\tilde{z} = 0$.
In quantum~theory, this relation holds in the rest
frame, i.e.~$\langle \tilde{\chi}^3(\tilde{t})\rangle = \langle \tilde{z}(\tilde{t}) \rangle\exp(H\tilde{t})$,~but
not, generically, in the non-rest frame. This plot shows that the expectation value of $\tilde{\chi}^3$ 
computed in the non-rest frame, $\langle \tilde{\chi}^3(t) \rangle$, is close to zero if $M \gg D \gg H$
holds. In this very case, the non-rest- and rest-frame calculations of $\tilde{\chi}^3$ agree with each other.
Right:~Numerical~calculation~of~$\langle \tilde{p}_z(t) \rangle$.~The expectation value of $\tilde{p}_z$ computed
in the non-rest frame tends to zero~if~$M \gg D \gg H$, which agrees with its expectation value
computed in the rest frame.
}\label{fig:tilde-position&momentum}
\end{figure}

Specifically, in the Minkowski-spacetime limit, i.e. $H \rightarrow 0$, we have
\beqa
\langle \tilde{z}(t) \rangle &\xrightarrow[H \,\rightarrow\, 0]{}&\frac{\sqrt{M^2+P^2}}{M}\,\langle z(t) \rangle - \frac{P}{M}\,t\,,
\eeqa
which vanishes if and only if $\langle z(t) \rangle = z(t)$. This equality approximately holds 
iff $M \gg D$, as shown in Sec.~\ref{sec:gwp-p}. Now, if we assume $H > 0$, then
$\langle \tilde{z}(t) \rangle$ can be expected~to~approach~zero if the packet
propagates along the classical geodesic~\eqref{eq:classical-trajectory}, i.e. $M \gg D \gg H$
must be satisfied. If otherwise, $\langle \tilde{z}(t) \rangle$ may be non-vanishing. This discrepancy
is illustrated on fig.~\ref{fig:tilde-position&momentum}, left.\footnote{The
integrand in~\eqref{eq:tilde-position}
turns out to be singular at $x=y=0$ and $z = (P/(\sqrt{M^2+P^2}-M)\pm\exp(-Ht))/H$ for all
$Ht \geq 0$. This singularity originates from the scale factor $\exp(H\tilde{t})$ which, if
expressed through~$t$~and~$\mathbf{x}$ by using~\eqref{eq:ds-to-rf}, vanishes at those
points. Therefore, the expectation value of
$\tilde{\chi}^3(t,\mathbf{x}) = \tilde{z}(t,\mathbf{x})\exp(H\tilde{t}(t,\mathbf{x}))$ instead
of $\tilde{z}(t,\mathbf{x})$ is shown in that figure. This problem is non-existent if those points
lie outside the wave-packet support, e.g., if $P/M = 1$, that is the case for a packet whose
support is inside the Hubble volume.}

The rest-frame four-momentum of the wave packet can also be calculated from the non-rest-frame
perspective:
\beqa\label{eq:tilde-momentum}
\langle \tilde{p}_\mu(\Sigma) \rangle &=& {\int_{\Sigma}}d\tilde{\Sigma}^\nu(\tilde{x})\,
\langle\tilde{\Theta}_{\mu\nu}(\tilde{x})\rangle
\;=\;
{\int_\Sigma}d\Sigma^\rho(x)\,\frac{\partial x^\lambda}{\partial \tilde{x}^\mu}\,
\langle\Theta_{\lambda\rho}(x)\rangle\,,
\eeqa
where $\langle\hat{\Theta}_{\mu\nu}(x)\rangle$ is a second-rank tensor, because
$\varphi_{X,P}(x)$ is a relativistic scalar. In Minkowski spacetime,
$\langle \tilde{p}_\mu(\Sigma) \rangle$ coincides with $\langle \tilde{p}_\mu(\tilde{\Sigma}) \rangle$
in the absence of external forces. This happens~to be the case, since the four-momentum
is then independent on Cauchy surfaces. In de-Sitter spacetime, their coincidence might
be expected if $M \gg D \gg H$ holds (see~fig.~\ref{fig:tilde-position&momentum}, right).

To summarise, the non-rest- and rest-frame computations of the position and momentum
expectation values are related like in classical physics if the wave packet behaves as a
point-like particle. According to the numerical calculations, this requires
the Compton~wavelength to be much smaller than the wave-packet localisation
size which, in turn, must be negligibly small relative to the curvature length. The former
is needed for the suppression of quantum features of the packet, while the later for it to be
oblivious to the global geometry. 

\newpage
\section{Discussion}

Elementary particles are described by wave packets in quantum theory. A wave packet in Minkowski
spacetime is usually constructed through the superposition of positive-frequency plane-wave solutions
of a given field equation~\cite{Itzykson&Zuber}. This wave packet can in turn be associated with an
asymptotic state used in the definition of $S$-matrix. But, the plane-wave solutions may exist only locally in
non-flat spacetimes. The basic question is then how to construct a wave packet
to describe a free elementary particle in the Universe.

In flat de-Sitter spacetime, one believes that the exact solution~\eqref{eq:a-modes-in} is
appropriate for the definition of elementary particles at past infinity, while the
solution~\eqref{eq:a-modes-out} is usually suggested for the description of particles at future infinity.
The superposition of each of~these~modes can certainly be used to construct Gaussian wave packets.
The three-momentum of these~packets are given by the Minkowski result~\eqref{eq:3-momentum}.
Hence, we should assume $M \gg D$ for each packet. In addition, we should assume
$D \gg H$, otherwise these wave packets have spatial support to be larger than the
cosmological extent of de-Sitter spacetime. In general, any wave~packet should be well localised
within the Hubble scale in order to describe an elementary particle. Repeating numerical
calculations made in the previous sections, we find these two Gaussian wave packets propagate
along a curve which approaches the geodesic~\eqref{eq:classical-trajectory} if $M \gg D \gtrsim H$.
In this case, the energy of the packets also approaches the classical result~\eqref{eq:cl-energy}.

Still, the adiabatic wave packets cannot be locally represented through the superposition of
plane waves and depend on coordinates used to parametrise the de-Sitter hyperboloid.~All
these mean that the adiabatic wave packets are locally described by phase factors which may differ
from $e^{-iM\tau}$, where $\tau$ is the proper time. In particular, their on-mass-shell phase factors depend
explicitly on the three-momentum. Specifically, taking the same initial conditions as in the previous section,
we find that that difference becomes more pronounced if we increase the ratio $P/M$ for fixed
$M/D = 10$ and $M/H = 100$. In fact,
expressing $t$ and $z$ through~the proper time $\tau$ and then considering $H\tau \ll 1$, we obtain
\bsubeqs
\beqa
a(t(\tau)) &\approx& 1 + \sqrt{1+\frac{P^2}{M^2}}\,H\tau\,,
\\[1mm]
z(\tau) &\approx& \frac{P}{M}\,\frac{\tau}{a(t(\tau))}\,.
\eeqa
\esubeqs
The adiabatic-wave-packets phase factors do reduce to $e^{-iM\tau}$ iff $M/H \gg 1$ and
$P/M \lesssim 1$. This is no longer the case if $P/M \gg 1$ and $PH\tau/M \gtrsim 1$ hold.
If we now assume that~$\tau \sim 1\,\text{s}$, $H = H_0 \sim 10^{-18}\,\text{s}^{-1}$ and
$M \sim 1\,\text{MeV}$, then the discrepancy should appear for $P \gtrsim 10^{15}\,\text{GeV}$.
This result turns out to be counter-intuitive, because high-energy physics
should not depend on the space-time curvature. This property of the adiabatic wave packets
might thereby lead to non-standard results for flavour oscillations and for the quantum interference
induced by gravity. However, $H_0 $ is way too small for any experimental tests of that and,
moreover, the observable Universe can be modelled by de-Sitter spacetime at
cosmological scales only.~It means this issue should be studied in Schwarzschild spacetime which
approximately describes the local geometry of Earth (see below for more details on this point).
\begin{figure}
\centering
\includegraphics[scale=0.9]{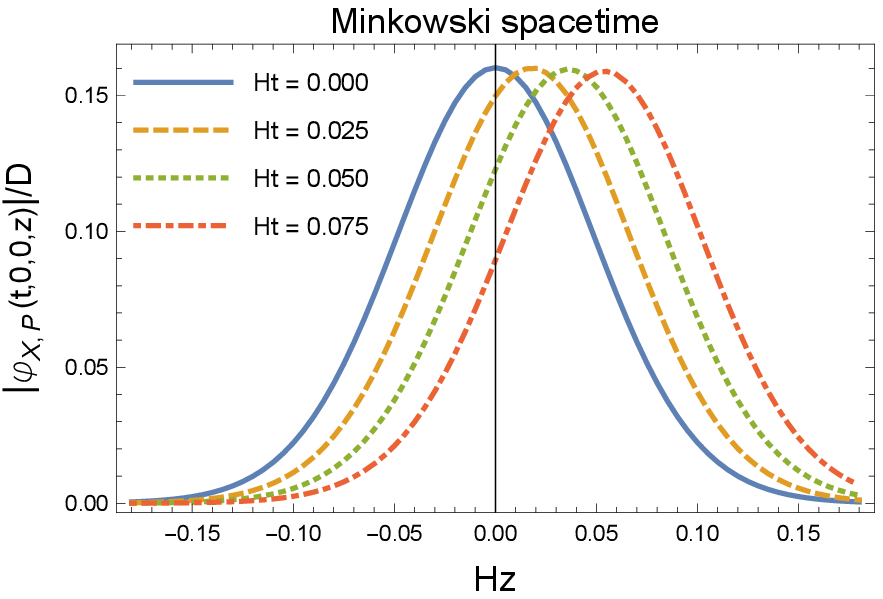}\hspace{2mm}
\includegraphics[scale=0.9]{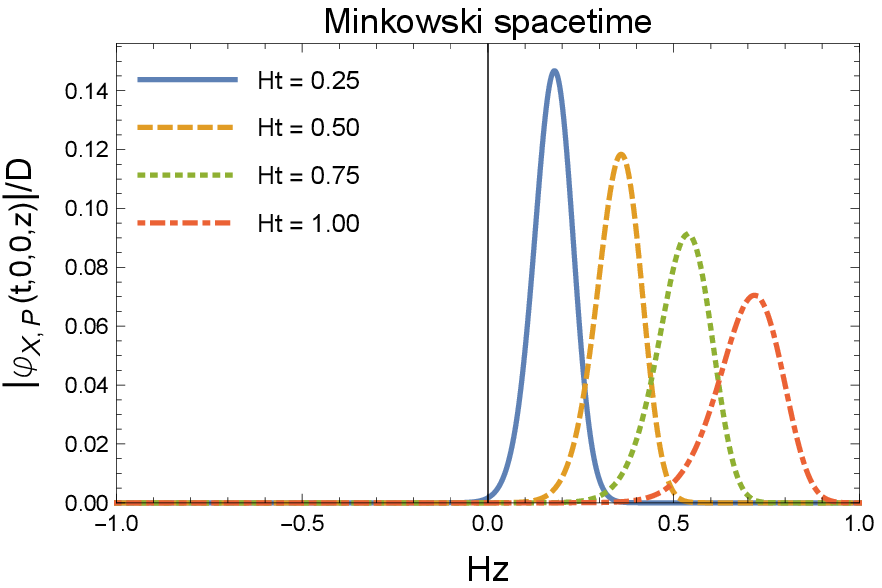}\vspace{4mm}
\includegraphics[scale=0.9]{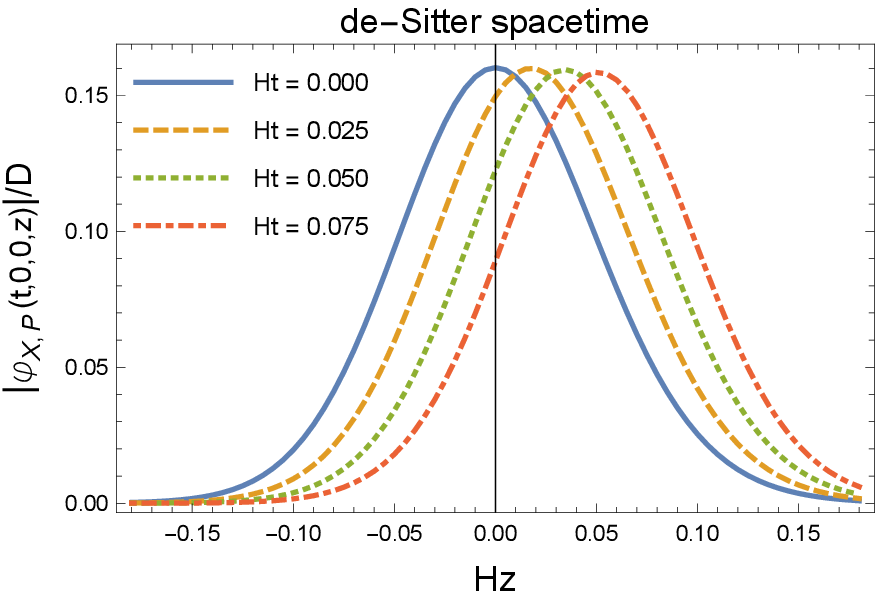}\hspace{2mm}
\includegraphics[scale=0.9]{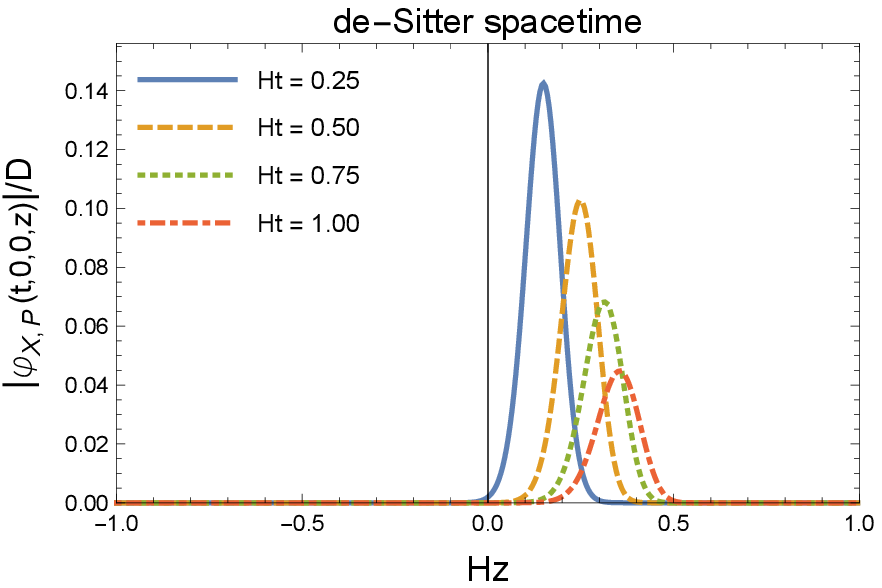}
\caption{The absolute value of the wave packet $\varphi_{X,P}(x)$ as a function of time in
Minkowski spacetime (top panel) and in de-Sitter spacetime (bottom panel), assuming that
$M/D = 10$, $M/H = 100 \gg 1$ and the initial conditions $X = 0$ with $P = (\sqrt{2}M,0,0,M)$.
Note, we make use of dimensionless~time and space coordinates, i.e. $Ht$ and $Hz$,
where, in Minkowski spacetime, $H$ is merely a dimensionful parameter, while, in de-Sitter spacetime,
$H$ is equal to its Hubble constant.
In de-Sitter spacetime, the wave
packet~\eqref{eq:gwf-ds} evolves as~\eqref{eq:gwp-m} in flat spacetime
for small values of $Ht$, in accordance with the equivalence principle. At later times,
$Ht \gtrsim 0.1$, the Minkowski and de-Sitter wave packets behave differently. One can also
easily see that the probability to find 
a scalar-field particle~on~the~classical geodesic is maximal in both cases.}\label{fig:mi-wp-vs-ds-wp}
\end{figure}

The main purpose of this article was to derive a wave packet which is a relativistic scalar
and locally reduces to the plane-wave superposition at any space-time point, no
matter~if~that point lies at past or future cosmic infinity or somewhere in between.
These basic properties are required for a wave packet to be appropriate for the description
of an elementary~particle. In particle physics, the Minkowski-spacetime
approximation of the Universe is underlying~for the definition of particles which are
related to the unitary and irreducible representations of the Minkowski-spacetime
isometry group~\cite{Weinberg}. Thus, those properties guarantee that~the~wave packet
locally corresponds to the irreducible representation of mass $M$ and spin $0$.
The main result of the article is that we have found that such a packet does exist in
de-Sitter~spacetime. This packet propagates over cosmological times like a point-like
particle of the same mass if $M \gg D \gg H$ (see also fig.~\ref{fig:mi-wp-vs-ds-wp}),
such that the wave-packet phase is characterised by $e^{-iM\tau}$,~as
expected in the semi-classical limit~\cite{Stodolsky}.
If otherwise, it propagates highly non-classically.

This wave-packet solution, $\varphi_{X,P}(x)$, gives rise to the particle-annihilation operator
\beqa
\hat{a}(\varphi_{X,P}) &=&
+ i{\int_\Sigma}d\Sigma^\mu(x)\,
\big(\overline{\varphi}_{X,P}(x)\nabla_{\mu}\hat{\Phi}(x)
-\hat{\Phi}(x)\nabla_{\mu}\overline{\varphi}_{X,P}(x)\big)\,.
\eeqa
This operator has two basic properties, namely it does not depend on the Cauchy
surface~$\Sigma$ and on the coordinates $x$ used to parametrise the de-Sitter
hyperboloid. The former property comes from, first, the absence of non-linear terms
in the scalar-field equation and, second, the localisation of $\varphi_{X,P}(x)$ on the
Cauchy surface $\Sigma$. The latter property is due to the covariant character of the
Klein-Gordon product and the wave-packet solution $\varphi_{X,P}(x)$. Thus,
$\hat{a}(\varphi_{X,P})$ defines a coordinate-independent quantum vacuum
($\hat{a}(\varphi_{X,P})|\Omega\rangle = 0$) in de-Sitter spacetime, 
while its Hermitian conjugate defines a covariant particle state
($|\varphi_{X,P}\rangle=\hat{a}^\dagger(\varphi_{X,P})|\Omega\rangle$).

The quantum state $|\Omega\rangle$ is a no-covariant-particle state which still may
be non-empty with respect to the de-Sitter particles which have been introduced in Sec.~\ref{sec:cpd-ds}.
To clarify this issue, the Bogolyubov coefficients need to be computed:
\bsubeqs
\beqa
\alpha_{X,P}(\mathbf{k}) &\equiv& -ie^{3Ht}{\int_t}d^3\mathbf{x}\,\big(
\varphi_{X,P}(x)\partial_t \bar{\varphi}_{\mathbf{k},-\infty}(x)
-\bar{\varphi}_{\mathbf{k},-\infty}(x)\partial_t\varphi_{X,P}(x)
\big)\,,
\\[0mm]
\beta_{X,P}(\mathbf{k}) &\equiv& -ie^{3Ht}{\int_t}d^3\mathbf{x}\,\big(
\bar{\varphi}_{X,P}(x)\partial_t \bar{\varphi}_{\mathbf{k},-\infty}(x)
-\bar{\varphi}_{\mathbf{k},-\infty}(x)\partial_t\bar{\varphi}_{X,P}(x)
\big)\,,
\eeqa
\esubeqs
which are time-independent due to the spatial localisation of $\varphi_{X,P}(x)$. Having used
the same initial conditions for $X$ and $P$ as in the previous section, we numerically find for
$\mathbf{k} = \mathbf{P}$ that $\alpha_{X,P}(\mathbf{k})$ is time-independent, while
$\beta_{X,P}(\mathbf{k})$ fluctuates with time and~$|\beta_{X,P}(\mathbf{k})/\alpha_{X,P}(\mathbf{k})| \lll 1$
decreases if an integration region increases. The same
calculations with $\varphi_{\mathbf{k},-\infty}(x)$ replaced~by $\varphi_{\mathbf{k},+\infty}(x)$
yield the Bogolyubov coefficients which are time-independent.
These observations might mean that $|\Omega\rangle$ is unitarily equivalent to the state
$|\text{dS}\rangle$. In any case, $\varphi_{X,P}(x)$
is related to the 2-point function in the de-Sitter~state, as shown in
Sec.~\ref{sec:ii-io}. Specifically, $\varphi_{X,P}(x)$ is proportional to that function if
$D \rightarrow \infty$. This turns out to be analogous to the Minkowski case, namely
the packet~\eqref{eq:gwp-m} is also proportional to the Minkowski
2-point function if the momentum variance of the wave packet is infinite. In both cases,
the proportionality coefficient is given by the normalisation factor.

The de-Sitter universe is not only one curved spacetime which is of physical interest from
the viewpoint of elementary particle physics. For example, black-hole spacetimes serve as a
non-trivial background for probing predictions of quantum theory. Field-equation solutions which are
commonly employed to define particles depend explicitly on global symmetries of a
given black-hole geometry~\cite{DeWitt-1975}. Still, the observable Universe can locally be
approximated by such a geometry only nearby a black hole this geometry is supposed to
describe. Thereby, black-hole global symmetries are local for the Universe. This circumstance poses
a question why those global symmetries should be ``preferred" with respect to local Poincar\'e
symmetry, taking into account that both are non-exact in the Universe. Since local Poincar\'e
symmetry is well known to play a crucial role in elementary particle physics~\cite{Weinberg},
one might actually need to re-consider those solutions which are employed to define elementary particles
in black-hole spacetimes. The reason is that those solutions like $\varphi_{\mathbf{k},-\infty}(x)$ and
$\varphi_{\mathbf{k},+\infty}(x)$ give rise to wave packets which cannot be everywhere represented
locally through the superposition of plane waves, as required by the equivalence
principle and collider physics. This~might~lead~to~novel results in quantum-black-hole
physics, such as~\cite{Emelyanov}.

This question
does not seem to be unanswerable nearby Earth, because its local curvature length is around
$10^{11}\,\text{m}$, which is 15 orders of magnitude smaller than that of the observable
Universe as a whole. This would \emph{naively} reduce the threshold momentum to around
$1\,\text{GeV}$, above which the standard mode solutions in Schwarzschild spacetime
may lead to the local-Lorentz-symmetry violation under certain experimental conditions. We
shall report on this study in one of our future publications.

\section*{
ACKNOWLEDGMENTS}

It is my pleasure to thank Prof.~O. Nachtmann for providing the reference~\cite{Nachtmann1968}.
I am also grateful to Martin Gabelmann for his support by working with the ITP computer cluster.

\end{document}